\documentclass[11 pt]{article}
\usepackage[mathscr]{eucal}
\usepackage{amssymb,latexsym}
\usepackage{verbatim}
\usepackage{amsmath}
\usepackage{amsthm}
\usepackage{enumerate}
\usepackage{authblk}
\usepackage{color}
\usepackage{multicol}
\usepackage{tikz}
\usepackage{xypic}
\usepackage{caption}
\usepackage{cite}
\usepackage{hyperref}
\usepackage[nottoc,notlot,notlof]{tocbibind}

\newtheorem{thm}{Theorem}[section]
\newtheorem{Def}[thm]{Definition}
\newtheorem{prop}[thm]{Proposition}

\hypersetup{
    colorlinks=true,
    linkcolor=black,
    filecolor=magenta,      
    urlcolor=cyan,
    citecolor=black,
}

\renewcommand\l{\lambda}

\newcommand\e{\varepsilon}
\renewcommand\b{\beta}

\renewcommand\l{\lambda}
\newcommand\g{\gamma}

\renewcommand\a{\alpha}

\newcommand\beq{\begin{equation}}
\newcommand\eeq{\end{equation}}
\newcommand\ben{\begin{enumerate}}
\newcommand\een{\end{enumerate}}
\newcommand\bit{\begin{itemize}}
\newcommand\eit{\end{itemize}}

\newcommand{\R}{\mathbb R}

\newcommand{\ov}{\overline}

\newcommand{\pd}{\partial}

\newcommand{\mc}{\mathcal}

\def\undertilde#1{\mathord{\vtop{\ialign{##\crcr
   $\hfil\displaystyle{#1}\hfil$\crcr\noalign{\kern1.5pt\nointerlineskip}
   $\hfil\tilde{}\hfil$\crcr\noalign{\kern1.5pt}}}}}

\usepackage{amsthm}

\newtheorem{innercustomgeneric}{\customgenericname}
\providecommand{\customgenericname}{}
\newcommand{\newcustomtheorem}[2]{%
  \newenvironment{#1}[1]
  {%
   \renewcommand\customgenericname{#2}%
   \renewcommand\theinnercustomgeneric{##1}%
   \innercustomgeneric
  }
  {\endinnercustomgeneric}
}

\newcustomtheorem{customthm}{Theorem}
\newcustomtheorem{customlemma}{Lemma}
\newcustomtheorem{customprop}{Proposition}
\newcustomtheorem{customcor}{Corollary}

\newcounter{mnotecount}

\title{A lower semicontinuous time separation function \linebreak for $C^0$ spacetimes}

\author{Eric Ling\footnote{el@math.ku.dk}}

\affil{Copenhagen Centre for Geometry and Topology (GeoTop),
\\
Department of Mathematical Sciences, University of Copenhagen, Denmark}

\begin{document}
\date{}
\maketitle
\vspace{.2in}

\begin{abstract}
The time separation function (or Lorentzian distance function) is a fundamental tool used in Lorentzian geometry. For smooth spacetimes it is known to be lower semicontinuous, and, in fact, continuous for globally hyperbolic spacetimes. Moreover, an axiom for Lorentzian length spaces -- a synthetic approach to Lorentzian geometry -- is the existence of a lower semicontinuous time separation function. Nevertheless, the usual time separation function is \emph{not} necessarily lower semicontinuous for $C^0$ spacetimes due to bubbling phenomena. In this paper, we introduce a class of curves called ``nearly timelike" and show that the time separation function for $C^0$ spacetimes is lower semicontinuous when defined with respect to nearly timelike curves. Moreover, this time separation function agrees with the usual one when the metric is smooth. Lastly, sufficient conditions are found guaranteeing the existence of nearly timelike maximizers between two points in a $C^0$ spacetime.

\end{abstract}


\vspace{.15in}

\tableofcontents

\newpage

\section{Introduction}

The time separation function (or Lorentzian distance function) is a fundamental tool used in Lorentzian geometry. For smooth spacetimes, it's known to be lower semicontinuous, and, in fact, continuous if the spacetime is globally hyperbolic \cite{ON}. In synthetic approaches to Lorentzian geometry, a lower semicontinuous time separation function even appears as an axiom, as in the Lorentzian length spaces of \cite{KS} or the recently introduced bounded Lorentzian-metric spaces of \cite{BLMS}.

However, for $C^0$ spacetimes, i.e., ones where the metric is only continuous, it's known that the time separation function -- as it's usually defined -- is \emph{not} lower semicontinuous due to bubbling phenomena. (Bubbling occurs when the causal future $J^+(p)$ is not contained in the closure of the timelike future $I^+(p)$ of a point $p$ in the spacetime.) The goal of this paper is to introduce a time separation function for $C^0$ spacetimes that is lower semicontinuous and agrees with the usual definition in the smooth setting. 

Let us briefly demonstrate what goes wrong in the continuous setting with the usual definition of the time separation function. If $p$ is the vertex of a bubbling region (e.g., the origin in \cite[Ex. 1.11]{ChrusGrant}) and $q$ is a point in the bubble, then there are causal curves with positive Lorentzian length from $p$ to $q$, but any neighborhood of $p$ contains points which are not in the causal past of $q$. The time separation between these points and $q$ is zero by definition, hence the time separation function is not lower semicontinuous. Another example appears before Theorem \ref{thm: lsc} below.

This paper is organized as follows. In section \ref{sec: preliminaries}, we review standard causal theory for $C^0$ spacetimes. In section \ref{sec: nearly timelike curves}, we introduce a new class of curves dubbed \emph{nearly timelike} and the relation: $q \in \mc{J}^+(p)$  if there is a nearly timelike curve from $p$ to $q$ or if $q = p$. It satisfies $I^+(p) \subset \mc{J}^+(p) \subset J^+(p)$. We define the time separation function with respect to nearly timelike curves instead of causal curves. Specifically, we define \[
\tau(p,q) \,=\, \sup \{L(\g) \mid \g \text{ is a nearly timelike curve from $p$ to $q$}\}
\]
whenever there is a nearly timelike curve from $p$ to $q$ 
and $\tau(p,q) = 0$ otherwise. Here $L(\g)$ is the Lorentzian length of $\g$. With this definition, we show that $\tau$ is lower semicontinuous in Theorem \ref{thm: lsc}. Moreover, $\tau$ satisfies the reverse triangle inequality and $\tau(p,q) > 0$ if and only if $q \in I^+(p)$. Furthermore, our definition agrees with the usual definition for the time separation function whenever the metric is smooth (locally Lipschitz is sufficient). At the end of section \ref{sec: nearly timelike curves}, we show how $C^0$ spacetimes fit into the framework of Lorentzian pre-length spaces. However, they do \emph{not} necessarily fit into the framework of Lorentzian length spaces since they are not locally causally closed: limit curves do not necessarily exist for nearly timelike curves. In section \ref{sec: nearly timelike maximizers}, we find sufficient conditions to prove a limit curve theorem for nearly timelike curves. We use this to prove the existence of \emph{nearly timelike maximizers} between two points under the aforementioned conditions. We end with a discussion and conclusion in section \ref{sec: conclusion}.

This work was motivated in part by investigating how $C^0$ spacetimes fit into the framework of Lorentzian (pre)-length spaces. These spaces were introduced in the influential work of Kungzinger and S{\"a}mann \cite{KS}. Since then numerous directions and generalizations have been explored,  see, for example,   \cite{cones_as_length_spaces, length_spaces_causal_hierarchy, time_fun_on_length_spaces, Lorentzian_analogue, null_distance_lorentzian_length_spaces, length_space_splitting, length_space_comparison, length_space_gluing, length_space_hyperbolic_angle, length_space_inextend, length_spaces_causal_completion, length_spaces_causal_ladder_gluing, length_spaces_Bonnet_Myers,
Flores_et_al, Cavalletti_Mondino, Cavalletti_Mondino_review}.

\medskip
\medskip

\section{$C^0$ spacetimes}\label{sec: preliminaries}

Treatments of causal theory in $C^0$ spacetimes can be found in \cite{Ling_causal_theory, Minguzzi_cone, ChrusGrant, Clemens_GH, Sorkin_Woolgar}.
We follow the conventions in \cite{Ling_causal_theory}, which we briefly review. A $C^k$ \emph{spacetime} is a smooth manifold $M$ (connected, Hausdorff, and second-countable) equipped with a $C^k$ Lorentzian metric $g$ and a time orientation induced by some $C^1$ timelike vector field $T$.\footnote{For most purposes, a $C^0$ timelike vector field is sufficient; the higher regularity is only important whenever we want the integral curves of $T$ to be unique. } (Our convention is that a nonzero vector $X$ is \emph{timelike, null,} or \emph{spacelike} if $g(X,X) < 0,\, = 0,\, > 0$, respectively.) If $X$ is either timelike or null, then $X$ is called \emph{causal}. A causal vector $X$ is \emph{future directed} if $g(X,T) < 0$ and \emph{past directed} if $g(X,T) > 0$. (Future-directed vectors become past directed with respect to the time orientation induced by $-T$. Consequently, definitions and theorems can be made with respect to the future, and the corresponding definitions and theorems for the past can be inferred.)

Fix a smooth complete Riemannian metric $h$ on $M$, and let $I \subset \R$ be an interval. A \emph{locally Lipschitz} curve $\g \colon I \to M$ is a continuous function such that for any compact $K \subset I$, there is a constant $C$ such that for any $a,b \in K$, we have $d_h\big(\g(a), \g(b)\big) \leq C|b-a|$, where $d_h$ is the Riemannian distance function associated with $h$. If $\g$ is locally Lipschitz with respect to some complete Riemannian metric $h$, then it's locally Lipschitz with respect to any other complete Riemannian metric \cite[Prop. A.5]{Ling_causal_theory}, so the choice of $h$ is nonessential. If $\g$ is locally Lipschitz, then the components $\g^\mu = x^\mu \circ \g$ in any coordinate system $x^\mu$ are differentiable almost everywhere and $(\g^\mu)' \in L^\infty_{\rm loc}$. 

 We will sometimes write $\g$ instead of $\g(I)$ for the image of $\g$. (This convention was also used in \cite{Ling_causal_theory}.) Also, if we say $\g$ is a curve from a point $p$ to another point $q$, then we mean there is a compact domain $[a,b]$ for $\gamma$ such that $\g(a) = p$ and $\g(b) = q$.

A \emph{causal curve}\footnote{Note that ``future directed" is implicit in the definition of causal and timelike curves. Therefore \emph{all} causal and timelike curves in this paper will be future directed. This convention was also used in \cite{Ling_causal_theory}. } is a locally Lipschitz curve $\g \colon I \to M$ such that $\g'$ is future directed causal almost everywhere. If a causal curve $\g$ also satisfies $g(\g', \g') < -\e$ almost everhywhere for some $\e > 0$, then $\g$ is called a \emph{timelike curve}. This class includes the piecewise $C^1$ timelike curves. To contrast, a causal curve $\g$ which simply satisfies $g(\g', \g') < 0$ almost everywhere is called an \emph{almost everywhere timelike curve}. The \emph{Lorentzian length} of a causal curve $\g \colon I \to M$ is $L(\g) = \int_I \sqrt{-g(\g', \g')}$. 

%


The \emph{$h$-arclength} of a locally Lipschitz curve $\g\colon I \to M$ is $L_h(\g) = \int_I \sqrt{h(\g', \g')}$, and $\g$ is \emph{parameterized by $h$-arclength} if $h(\g',\g') = 1$ almost everywhere.    Causal curves can always be reparameterized by $h$-arclength \cite[Prop. 2.14]{Ling_causal_theory}. If $\g \colon (a,b) \to M$ is a causal curve parameterized by $h$-arclength, then $\g$ is inextendible as a causal curve if and only if $(a,b) = \R$; moreover, if say $b < \infty$, then $\g$ can be made future inextendible by, for example,  concatenating it with the maximal integral curve of a timelike vector field.

The \emph{causal future} of a point $p \in M$, denoted by $J^+(p)$, is the union of $p$ itself together with the set of points $q \in M$ which can be reached by a causal curve starting from $p$. That is, $q \in J^+(p)$ if and only if there is a causal curve $\g$ from $p$ to $q$ or if $q = p$. The \emph{timelike future} of a point $p \in M$, denoted by $I^+(p)$, is simply the set of points $q \in M$ that can be reached by a timelike curve starting from $p$. The \emph{causal past} $J^-(p)$ and \emph{timelike past} $I^-(p)$ are defined time-dually. $I^\pm(p)$ are open sets \cite[Thm. 2.12]{Ling_causal_theory}. We will occasionally  write $p \ll q$ to denote $q \in I^+(p)$. 

\medskip
\medskip

\noindent{\bf Remark 2.1.} Other references use $I^+(p)$ to denote the timelike future with respect to almost everywhere timelike curves; we use the notation $I^+_{\rm a.e.}(p)$ for this, see the appendix. Although this choice may seem more natural, the drawback is that $I^+_{\rm a.e.}(p)$ is not necessarily open \cite{future_not_open}. In the appendix, we review the different definitions of timelike curves (e.g., piecewise $C^1$ and locally uniformly timelike) and show that their corresponding timelike futures all coincide (see Proposition \ref{prop: timelike futures}), at least for those whose timelike futures form open sets.

\medskip
\medskip

If $I^+\big(J^+(p)\big) = I^+(p)$ and its time-dual statement $I^-\big(J^-(p)\big) = I^-(p)$ hold for all $p \in M$, then $(M,g)$ is said to satisfy the \emph{push-up property for $J$}. If the metric $g$ is locally Lipschitz (i.e., its components in any coordinate system are locally Lipschitz functions), then it is known that $(M,g)$ satisfies the push-up property for $J$ (see \cite[Lem. 1.15]{ChrusGrant} or \cite[Thm. 4.15]{Ling_causal_theory}). However, it can fail below this regularity; in this case, pathologies like \emph{causal bubbles} can form (i.e., ``future bubbling" regions where $\text{int}\big[J^+(p)\big] \setminus I^+(p)$) is nonempty). Examples and discussions of spacetimes with bubbling can be found in \cite{ChrusGrant, Ling_causal_theory, Leonardo_Soultanis, SbierskiUniqueness, future_not_open, Clemens_Steinbauer}. 

The next proposition gives equivalent formulations of the push-up property for $J$. It motivates the definition of ``nearly timelike curves" given in the next section.

\medskip

\begin{customprop}{2.2}\label{prop: push-up}
For a $C^0$ spacetime $(M,g)$, the following are equivalent.
\begin{itemize}
\item[\emph{(1)}] $I^+\big(J^+(p)\big) = I^+(p)$. 
\item[\emph{(2)}] $\emph{int}\big[J^+(p)\big] = I^+(p)$. 
\item[\emph{(3)}] $J^+(p) \subset \ov{I^+(p)}$. 
\end{itemize}
\end{customprop}

\proof \;

 (1) implies (2): Since $I^+(p)$ is open and contained in $J^+(p)$,  it's also contained in the interior of $J^+(p)$. For the reverse inclusion, fix $q \in \text{int}\big[J^+(p)\big]$. $I^-(q)$ meets every neighborhood of $q$; hence it meets $\text{int}\big[J^+(p)\big]$ at some point $r$. Since $r$ is in the interior, we can assume $r \neq p$. Therefore there is a causal curve from $p$ to $r$ and a timelike curve from $r$ to $q$.  Therefore $q \in I^+(p)$ by (1).

 (2) implies (3):  Fix $q \in J^+(p)$. Since $I^+(q)$ is an open set contained in $\text{int}\big[J^+(p)\big]$, we have $I^+(q) \subset I^+(p)$ by assumption. Thus $q \in \ov{I^+(p)}$ since $I^+(q)$ meets any neighborhood $U$ of $q$.

(3) implies (1): Clearly $I^+(p) \subset I^+\big(J^+(p)\big)$. For the reverse inclusion, fix $q \in J^+(p)$ and $r \in I^+(q)$. Since $I^-(r)$ is an open set containing $q$ and $q \in \ov{I^+(p)}$, we have $I^-(r)$ meets $I^+(p)$. Therefore $r \in I^+(p)$. 
\qed

\medskip
\medskip

Suppose the push-up property for $J$ holds on a $C^0$ spacetime $(M,g)$. From the previous proposition and its time-dual version, it follows that for any causal curve $\g \colon [a,b] \to M$, we have 
\[
\g(t) \,\in\, \ov{I^+\big(\g(s)\big)} \quad \text{ and } \quad \g(s) \,\in\, \ov{I^-\big(\g(t)\big)}
\]
for all $s < t$ in $[a,b]$. It is this property we use to define ``nearly timelike curves" for $C^0$ spacetimes in the next section.

\section{Nearly timelike curves}\label{sec: nearly timelike curves}

 In this section, we introduce a relation $\mc{J}$ between $I$ and $J$ based on causal curves which never enter bubbling regions. Formally:

\medskip
\medskip

\begin{Def}\label{def: nearly timelike}
\emph{
Let $(M,g)$ be a $C^0$ spacetime. }
\begin{itemize}

\item[$\bullet$] \emph{Let $\g \colon [a,b] \to M$ be a causal curve. We call $\g$ a \emph{a nearly timelike curve} if
\[
\g(t) \,\in\, \ov{I^+\big(\g(s)\big)} \quad \text{ and } \quad \g(s) \,\in\, \ov{I^-\big(\g(t)\big)}
\]
for all $s < t$ in $[a,b]$. Clearly, timelike curves are nearly timelike. Moreover, the restriction of a nearly timelike curve is nearly timelike, i.e., if $\g \colon [a,b] \to M$ is a nearly timelike curve, then $\g|_{[c,d]}$ is a nearly timelike curve for any $[c,d] \subset [a,b]$.
}

\item[$\bullet$] \emph{ We define the \emph{nearly timelike future} of a point $p \in M$ as the set
\[
\mathcal{J}^+(p) \,=\, \{q \in M \mid \text{there is a nearly timelike curve from $p$ to $q$} \} \cup \{p\}.
\]
The \emph{nearly timelike past} $\mc{J}^-(p)$ is defined time-dually. Evidently, 
\[
q \in \mc{J}^+(p) \:\: \Longleftrightarrow \:\: p \in \mc{J}^-(q) \quad \quad \text{ and } \quad \quad I^+(p) \subset \mc{J}^+(p) \subset J^+(p).
\] }

\item[$\bullet$] \emph{The \emph{time separation function} $\tau \colon M \times M \to [0,\infty]$ will be defined with respect to nearly timelike curves. More precisely, if there is a nearly timelike curve from $p$ to $q$, we define
\[
\tau(p,q) \,=\, \sup \{L(\g) \mid \g \text{ is a nearly timelike curve from $p$ to $q$}\},
\]
and $\tau(p,q) = 0$ otherwise.
} 
\end{itemize}
\end{Def}

\medskip

Let $\tau_{\rm causal}$ denote the usual time separation function as it's normally defined \cite{ON}. The only difference between $\tau$ and $\tau_{\rm causal}$ is that the latter is defined with ``causal" curves, while the former is defined with ``nearly timelike" curves. Clearly $\tau \leq \tau_{\rm causal}$. They are equal whenever the push-up property for $J$ holds for a $C^0$ spacetime $(M,g)$; this follows since, in this case, $J^+(p) = \mc{J}^+(p)$ for all $p \in M$ via Proposition \ref{prop: push-up} in the previous section. The downside of using $\tau_{\rm causal}$ instead of $\tau$ is that the former is not necessarily lower semicontinuous when the spacetime does not satisfy the push-up property. This is demonstrated in the example below.
 
 \medskip
 \medskip
 
\noindent{\bf Example 3.2 (Garc{\'i}a-Heveling-Soultanis spacetime \cite{Leonardo_Soultanis}).} We show that $\tau_{\rm causal}$ is not necessarily lower semicontinuous. Let $(M,g)$ denote the bubbling $C^0$ spacetime in \cite{Leonardo_Soultanis}:
\[
M \,=\, \R^2 \quad \text{ and } \quad g \,=\, -dt^2 + \rho(t,x) dx^2, \quad \text{ where } \quad \rho(t,x) = 1 + \sqrt{(t-|x|)_+}.
\]
Note that the metric below $t = |x|$ is just the Minkowski metric. Let $p = (0,0)$ denote the origin. Fix a point $q \in \text{int}\big[J^+(p)\big]\setminus I^+(p)$, i.e., $q$ lies in the interior of the bubble. There are causal curves from $p$ to $q$ with positive Lorentzian length, hence $\tau_{\rm causal}(p,q) > 0$. Any neighborhood $U$ of $p$ contains points $p' \in U$ such that $p' \notin J^-(q)$ (e.g., take $p' \in I^+(p) \cap U$); therefore $\tau_{\rm causal}(p', q) = 0$, which implies $\tau_{\rm causal}$ is not lower semicontinuous. See Figure \ref{fig: 1}.

\begin{figure}[h]
\[
\begin{tikzpicture}[scale = 1]

\draw [thick] (0,-2) -- (4.1,2.1);
\draw [thick] (0,-2) -- (-4.1,2.1);

\draw [<->,thick] (0,-3.5) -- (0,2.35);
\draw [<->,thick] (-4.5,-2) -- (4.5,-2);

\draw (-.35,2.5) node [scale = .85] {$t$};
\draw (4.75, -2.25) node [scale = .85] {$x$};

\draw (1.75,3.15) node [scale = .85,white]{\small{A Milne-like spacetime}};

\draw [->] [thick] (-3.75,-0.5) arc [start angle=-90, end angle=-30, radius=40pt];
\draw (-4.5,-0.5) node [scale = .85] {$\pd J^+(p)$};

\draw [->] [thick] (-2,2.5) arc [start angle=10, end angle=-50, radius=35pt];
\draw (-2,2.9) node [scale = .85] {$\pd I^+(p)$};

\draw [thick, blue] (0,-2) .. controls (2.5,1) .. (3,2.1);
\draw [thick, blue] (0,-2) .. controls (-2.5,1) .. (-3,2.1);



\draw [thick, red] (0,-2)  -- (3,1);
\draw [thick, red] (3,1) -- (3.25, 1.75);

\node [scale = .50] [circle, draw, fill = black] at (0,-2)  {};
\draw (-.3,-2.3) node [scale = .85] {\small{$p$}};

\node [scale = .50] [circle, draw, fill = black] at (3.25,1.75)  {};
\draw (3.45,2.1) node [scale = .85] {\small{$q$}};

\node [scale = .50] [circle, draw, fill = black] at (0,-1.5)  {};
\draw (-.25,-1.15) node [scale = .85] {\small{$p'$}};

\end{tikzpicture}
\]
\caption{\small{In the Garc{\'i}a-Heveling-Soultanis example \cite{Leonardo_Soultanis}, there are causal curves from $p$ to $q$ with positive Lorentzian length such as the piecewise red curve in the figure. Hence $\tau(p,q) > 0$. However, there are points $p'$ arbitrarily close to $p$ with $p' \notin J^-(q)$; therefore $\tau_{\rm causal}(p', q) = 0$ by definition. Thus $\tau_{\rm causal}$ is not lower semicontinuous for this example.}}
\label{fig: 1}
\end{figure}

\medskip
\medskip

Although $\tau_{\rm causal}$ is not necessarily lower semicontinuous for $C^0$ spacetimes, the next theorem shows that $\tau$ from Definition \ref{def: nearly timelike} is.

\medskip

\begin{customthm}{3.3}\label{thm: lsc} The following hold for a $C^0$ spacetime $(M,g)$.
\begin{itemize}
\item[\emph{(1)}]  $I^+\big(\mc{J}^+(p)\big) = I^+(p)$ and $I^-\big(\mc{J}^-(p)\big) = I^-(p)$ for all $p \in M$.
\item[\emph{(2)}]$r \in \mc{J}^+(p)$ whenever $r \in \mc{J}^+(q)$ and $q \in \mc{J}^+(p)$.
\item[\emph{(3)}] $\tau(p,r) \geq \tau(p,q) + \tau(q,r)$ whenever $r \in \mc{J}^+(q)$ and $q \in \mc{J}^+(p)$.
\item[\emph{(4)}] $\tau(p,q) > 0$ if and only if $q \in I^+(p)$.
\item[\emph{(5)}] $\tau$ is lower semicontinuous.
\end{itemize}
\end{customthm}

\medskip

\noindent{\bf Remark 3.4.} (1) will be referred to as the \emph{push-up property for $\mc{J}$.} Unlike the usual push-up property for $J$, the push-up property for $\mc{J}$ \emph{always} holds for $C^0$ spacetimes. (2) shows that the relation $\mc{J}$ is transitive. (3) is known as the \emph{reverse triangle inequality}. (4) and (5) are axioms in the definition of a Lorentzian pre-length space \cite{KS}; after the proof of the theorem, we show how $C^0$ spacetimes fit into the framework of Lorentzian pre-length spaces.

\medskip

\newpage

\proof
\:

\begin{itemize}
\item[(1)]  Clearly $I^+(p) \subset I^+\big(\mc{J}^+(p)\big)$. For the reverse inclusion, fix $q \in \mc{J}^+(p)$ and $r \in I^+(q)$.  $I^-(r)$ is an open set containing $q$; hence $I^-(r)$ meets $I^+(p)$ since $q \in \ov{I^+(p)}$. The time-dual statement holds since the definition of a nearly timelike curve is time-symmetric.

\item[(2)] If $r = q$ or $q = p$, then the result is trivial, so suppose $r \neq q$ and $q \neq p$. Let $\a \colon[0,1]\to M$ and $\beta\colon[1,2] \to M$ be nearly timelike curves from $p$ to $q$ and $q$ to $r$, respectively. Let $\g \colon [0,2] \to M$ be the concatenation of $\a$ and $\b$. We show that $\g$ is a nearly timelike curve. Fix $s < t$ in $[0,2]$. The cases $t \leq 1$ or $s \geq 1$ are trivial to check since $\g(t)$ equals $\a(t)$ or $\beta(t)$, respectively, in these cases. Assume the remaining case: $s < 1$ and $t > 1$. Let $U$ be any neighborhood of $\g(t)$. Then $U$ intersects $I^+(q)$ since $\g(t) = \beta(t)$. Therefore there is a timelike curve from $q$ to some $x \in U$. Hence $x \in I^+\big(\g(s)\big)$ by the push-up property for $\mc{J}$, hence $\g(t) \in \ov{I^+\big(\g(s)\big)}$. The time-dual of this argument gives $\g(s) \in \ov{I^-\big(\g(t)\big)}$.


\item[(3)] There are four cases to consider. 

\underline{Case 1}: There is a nearly timelike curve from $p$ to $q$ and one from $q$ to $r$. Fix $\e > 0$. There is a nearly timelike curve $\a$ from $p$ to $q$ such that $\tau(p,q) \leq L(\a) + \e$. Likewise there is a nearly timelike curve $\b$ from $q$ to $r$ with $\tau(q,r) \leq L(\b) + \e$. If $\g$ denotes the concatenation of $\a$ and $\b$ (which is nearly timelike by (2)), then 
\[
\tau(p,r) \,\geq\, L(\g) \,=\, L(\a) + L(\b) \,\geq\, \tau(p,q) + \tau(q,r) - 2\e.
\]
Since $\e$ was arbitrary, the result follows.

\underline{Case 2}: There is a nearly timelike curve from $p$ to $q$ but none from $q$ to $r$. In this case, we must have $q = r$ and $\tau(q,q) = 0$. Therefore the reverse triangle inequality $\tau(p,r) \geq \tau(p,q) + \tau(q,r)$ reduces to $\tau(p,q) \geq \tau(p,q)$, which is clearly true.

\underline{Case 3}: There is no nearly timelike curve from $p$ to $q$, but there is one from $q$ to $r$. This case is similar to case 2.

\underline{Case 4}: There is no nearly timelike curve from $p$ to $q$ and also none from $q$ to $r$. In this case, we have $p = q = r$ and $\tau(p,p) = 0$. The reverse triangle inequality reduces to $0 \geq 0$, which is clearly true.

\item[(4)] Clearly $\tau(p,q) > 0$ whenever $q \in I^+(p)$. Conversely, suppose $\tau(p,q) > 0$. Then there is a nearly timelike curve $\g \colon [a,b] \to M$ from $p$ to $q$ with $L(\g) > 0$. Therefore there is a $t_0 \in (a,b)$ such that $\g'(t_0)$ is future-directed timelike. Without loss of generality, we can assume $t_0 = 0$ and $\g'(0)$ is unit, i.e., $g\big(\g'(0), \g'(0)\big) = -1$. Fix $\e > 0$. From \cite[Lem. 2.9]{Ling_causal_theory}, there is a coordinate neighborhood $U$ with coordinates $x^\mu$ around $\g(0)$ such that $\pd_0 = \g'(0)$. Set $\g^\mu = x^\mu \circ \g$. By definition of the derivative, there is a $\delta > 0$ such that $0 < t <\delta$ implies $|\g^0(t)/t - 1| < \e$ and $|\g^i(t)/t| < \e$ for all $i = 1, \dotsc, n$ (where $n +1$ is the dimension of the spacetime). Hence, for these $t$, we have $\g^0(t)/|\g^i(t)| > \tfrac{1-\e}{\e}$. 
Therefore, by choosing $\e$ small enough, we can find times $0 < t_1 < t_2 < \delta$, sufficiently close to $0$, such that $\g(t_1) \in I^+_{\eta^{\e}}\big(\g(0), U\big)$ and $\g(t_2) \in I^+_{\eta^\e}\big(\g(t_1), U\big)$, where $\eta^\e$ is the narrow Minkowskian metric on $U$ defined by $\eta^\e = -\frac{1-\e}{1+\e}(dx^0)^2 + \delta_{ij}dx^idx^j$. (This is clear geometrically but a rigorous argument can be given using Lemma 2.11(1) in\cite{Ling_causal_theory}). Then Lemma 2.9(5) in \cite{Ling_causal_theory} implies $\g(t_1) \in I^+\big(\g(0)\big)$ and $\g(t_2) \in I^+\big(\g(t_1)\big)$. See Figure \ref{fig: 2}.  Then there is a nearly timelike curve from $p$ to $\g(0)$ and a timelike curve from $\g(0)$ to $\g(t_1)$. Thus $\g(t_1) \in I^+(p)$ by (1), i.e., by an application of the push-up property for $\mc{J}$. Similarly, there is a timelike curve from $\g(t_1)$ to $\g(t_2)$ and a nearly timelike curve from $\g(t_2)$ to $q$; thus $\g(t_1) \in I^-(q)$ by an application of the time-dual push-up property for $\mc{J}$. Therefore there is a timelike curve from $p$ to $\g(t_1)$ and one from $\g(t_1)$ to $q$, hence $q \in I^+(p)$.

\begin{figure}[h]
\[
\begin{tikzpicture}[scale = 1]


\draw [<->,thick] (0,-3.5) -- (0,2.35);
\draw [<->,thick] (-4.5,-2) -- (4.5,-2);

\draw (-.35,2.5) node [scale = .85] {$x^0$};
\draw (4.75, -2.25) node [scale = .85] {$x^i$};

\draw [thick, blue] (0,-2) .. controls (0.2,1) .. (1,2.1);

\node [scale = .50] [circle, draw, fill = black] at (0.22,.5) {};
\draw (.74,.4) node [scale = .85] {\small{$\g(t_1)$}};

\node [scale = .50] [circle, draw, fill = black] at (1,2.1) {};
\draw (1.5,2) node [scale = .85] {\small{$\g(t_2)$}};

\node [scale = .50] [circle, draw, fill = black] at (0,-2)  {};
\draw (-.3,-2.3) node [scale = .85] {\small{$\g(0)$}};

\end{tikzpicture}
\]
\caption{\small{The proof of (4). $\g'(0)$ is future-directed timelike. Therefore, in a small Minkowskian neighborhood centered at $\g(0)$, we can find times $t_1$ and $t_2$ such that $\g(0) \ll \g(t_1) \ll \g(t_2)$.}}
\label{fig: 2}
\end{figure}

\item[(5)] Fix $(p_0, q_0) \in M \times M$ and $t < \tau(p_0, q_0)$. We want to show that there is a neighborhood of $(p_0, q_0)$ such that $\tau(p,q) > t$ for all $(p,q)$ in this neighborhood. If $\tau(p_0, q_0) = 0$, then $M \times M$ is such a neighborhood. Now assume $\tau(p_0, q_0) \in (0,\infty)$. Set $\e = \tau(p_0, q_0) - t.$ Let $\g \colon [0,1] \to M$ be a nearly timelike curve from $p_0$ to $q_0$ with $L(\g) > \tau(p_0, q_0) - \frac{\e}{2}$. Pick $0 < a < b < 1$ such that $0 < L(\g|_{[0,a]}) < \frac{\e}{4}$ and $0 < L(\g|_{[b,1]}) < \frac{\e}{4}$. Set $p' = \g(a)$ and $q' = \g(b)$. Since the restriction of a nearly timelike curve is nearly timelike, (4) implies that $p' \in I^+(p_0)$ and $q' \in I^-(q_0)$. Put $U = I^-(p')$ and $V = I^+(q')$. We show that $U \times V$ is the desired neighborhood. Indeed, for any $(p,q) \in U \times V$, the reverse triangle inequality implies
\begin{align*}
\tau(p,q) \,&\geq\, \tau(p, p') + \tau(p', q') + \tau(q', q) 
\\
&>\, \tau(p', q')
\\
&\geq\, L(\g|_{[a,b]})
\\
&=\, L(\g) - L(\g|_{[0,a]}) - L(\g|_{[b,1]} )
\\
&>\, \tau(p_0, q_0) - \e 
\\
&=\, t.
\end{align*} 
Lastly, if $\tau(p_0, q_0) = \infty$, then a similar argument as above yields the desired neighborhood.   \qed
\end{itemize}

\medskip

\noindent{\bf Remark 3.5.} The proof of (4) in the previous theorem has the following interesting consequence: Let $(M,g)$ be a $C^0$ spacetime and let $\g \colon I \to M$ be a nearly timelike curve from $p$ to $q$. If $q \notin I^+(p)$, then $\g$ does not contain a single timelike tangent (cf. Corollary 14.5 in \cite{ON}).

\medskip
\medskip

We use the previous theorem to show how $C^0$ spacetimes fit into the framework of Lorentzian pre-length spaces. A \emph{Lorentzian pre-length space} is a quintuple $(X,d, \ll, \leq, \tau)$ satisfying the following four axioms \cite{KS}.

\medskip
\medskip

\noindent\underline{Axioms for a Lorentzian pre-length space $(X,d,\ll, \leq, \tau)$}:
\begin{itemize}
\item[1.] $(X,d)$ is a metric space,
\item[2.] $\leq$ is a reflexive and transitive relation,
\item[3.] $\ll$ is a transitive relation contained in $\leq$,
\item[4.] $\tau \colon X \times X \to [0,\infty]$ is a lower semicontinuous map satisfying
\begin{itemize}
\item[(a)] $\tau(x,y) = 0$ if $x \nleq y$,
\item[(b)] $\tau(x,y) > 0$ if and only if $x \ll y$,
\item[(c)] $\tau(x,z) \geq \tau(x,y) + \tau(y,z)$ whenever $x \leq y \leq z$.
\end{itemize}
\end{itemize}
\medskip

\begin{customcor}{3.6}\label{cor: cont spacetimes are LpreLS}
Let $(M,g)$ be a $C^0$ spacetime with a complete Riemannian metric $h$ on $M$. Define the relations $\ll$ and $\leq$ via 
\[
p \,\ll\, q \quad \text{iff} \quad q \in I^+(p) \quad \quad \text{ and } \quad \quad p \,\leq\, q \quad \text{iff} \quad q \in \mc{J}^+(p).
\]
Let $\tau$ be the time separation function introduced in Definition \ref{def: nearly timelike}. Then $(M, d_h, \ll, \leq, \tau)$ is a Lorentzian pre-length space.
\end{customcor}

\proof
We verify the axioms of a Lorentzian pre-length space:
\begin{itemize}
\item[1.] $(M,d_h)$ is a metric space. In fact it's a complete metric space.
\item[2.] $\leq$ is reflexive since $p \in \mc{J}^+(p)$ by definition. It's transitive by Theorem \ref{thm: lsc}(2).
\item[3.] That $\ll$ is transitive follows immediately from the definition of a timelike curve. That $\ll$ is contained in $\leq$ means, by definition, that $p \ll q$ implies $p \leq q$. This holds since $I^+(p) \subset \mc{J}^+(p)$.
\item[4.] $\tau$ is lower semicontinuous by Theorem \ref{thm: lsc}(5). 
\begin{itemize}
\item[(a)] If $p \nleq q$, then there is no nearly timelike curve from $p$ to $q$. Therefore $\tau(p,q) = 0$ by definition.
\item[(b)] This follows from Theorem \ref{thm: lsc}(4).
\item[(c)] This follows from Theorem \ref{thm: lsc}(3). \qed
\end{itemize} 
\end{itemize}

\medskip
\medskip

We end this section with some comments on Lorentzian length spaces. A Lorentzian pre-length space $(X,d,\ll, \leq, \tau)$ is called \emph{locally causally closed} if for each $x \in X$, there is a neighborhood $U$ of $x$ such that if $p_n, q_n \in U$ are sequences converging to $p \in \ov{U}$ and $q\in \ov{U}$, respectively, with $p_n \leq q_n$ for all $n$, then $p \leq q$, see \cite[Def. 3.4]{KS}. See also the ``corrected" definition, \emph{weakly locally causally closed} in \cite[Def. 2.16]{length_spaces_causal_hierarchy}.
One of the requirements for a Lorentzian pre-length space to be a Lorentzian length space is that $(X,d, \ll, \leq, \tau)$ is locally causally closed, see \cite[Def. 3.22]{KS}. The following example shows that the Lorentzian pre-length space for a $C^0$ spacetime $(M,g)$, as defined in Corollary \ref{cor: cont spacetimes are LpreLS}, is not necessarily (weakly) locally causally closed even when $(M,g)$ is globally hyperbolic. 

\medskip
\medskip

\noindent{\bf Example 3.7.} Let $(M,g)$ be  the Garc{\'i}a-Heveling-Soultanis $C^0$ spacetime \cite{Leonardo_Soultanis}. We show that $(M,d_h, \ll, \leq, \tau)$ is not locally causally closed. Let $(t,x)$ be the natural coordinates on $M = \R^2$. Let $p = (0,0)$ denote the origin. For any $\e > 0$, let $B_{2\e}$ denote the usual open ball with radius $2\e$ centered at $p$. (By ``usual" we mean defined with with respect to the Euclidean metric $h = \delta$ on $M = \R^2$.) For $n = 1, 2, \dotsc$, set the points $p_n = (-\e/n, 0)$ on the negative $t$-axis. For all $n$, set $q_n = q = (\e, \e)$. Since the spacetime is isometric to Minkowski spacetime for points $t \leq |x|$, the straight line $\g_n$ joining $p_n$ to $q_n$ is timelike and lies entirely in $B_{2\e}$. Therefore $p_n \leq q_n$. However, $p_n \to p$ and $q_n \to q$ but $p \nleq q$ since there is no nearly timelike curve joining $p$ to $q$ (see \cite[Prop. 2.1]{Leonardo_Soultanis}). See Figure \ref{fig: 3}. Thus the corresponding Lorentzian pre-length space $(M,d_\delta, \ll, \leq, \tau)$ is not locally causally closed. (To show that it's not weakly locally causally closed, one makes use of Proposition 5.9 in \cite{KS}.)

\begin{figure}[h]
\[
\begin{tikzpicture}[scale = 1]

\draw [thick] (0,-2) -- (4.1,2.1);
\draw [thick] (0,-2) -- (-4.1,2.1);

\draw [<->,thick] (0,-4.75) -- (0,2.35);
\draw [<->,thick] (-4.5,-2) -- (4.5,-2);

\draw (-.35,2.5) node [scale = .85] {$t$};
\draw (4.75, -2.25) node [scale = .85] {$x$};

\draw [red, thick] (0,-2) -- (2,0);

\draw [->] [thick] (-3.75,-0.5) arc [start angle=-90, end angle=-30, radius=40pt];
\draw (-4.5,-0.5) node [scale = .85] {$\pd J^+(p)$};

\draw [->] [thick] (-2,2.5) arc [start angle=10, end angle=-50, radius=35pt];
\draw (-2,2.9) node [scale = .85] {$\pd I^+(p)$};

\draw [thick, blue] (0,-2) .. controls (2.5,1) .. (3,2.1);
\draw [thick, blue] (0,-2) .. controls (-2.5,1) .. (-3,2.1);


\node [scale = .50] [circle, draw, fill = black] at (0,-2)  {};
\draw (-.3,-2.3) node [scale = .85] {\small{$p$}};

\node [scale = .50] [circle, draw, fill = black] at (2,0)  {};
\draw (2.65,-.3) node [scale = .85] {\small{$q = q_n$}};

\node [scale = .50] [circle, draw, fill = black] at (0,-2.35)  {};
\node [scale = .50] [circle, draw, fill = black] at (0,-2.8)  {};
\node [scale = .50] [circle, draw, fill = black] at (0,-3.3)  {};

\draw (-.25,-3.5) node [scale = .85] {\small{$p_n$}};

\draw [thick] (0,-2.35) -- (2,0);
\draw [thick] (0,-2.8) -- (2,0);
\draw [thick] (0,-3.3) -- (2,0);

\draw (.75,-2.75) node [scale = .85] {\small{$\g_n$}};

\end{tikzpicture}
\]
\caption{\small{The Garc{\'i}a-Heveling-Soultanis spacetime \cite{Leonardo_Soultanis} is not (weakly) locally causally closed since the timelike curves $\g_n$ approach a causal curve (in red) which is not nearly timelike. }}
\label{fig: 3}
\end{figure}

\medskip

The previous example shows that, in general, a limit curve argument will not hold for nearly timelike curves. However, one does exist if additional assumptions are imposed, see Lemma \ref{lem: limit curve} in the next section.

\section{Nearly timelike maximizers}\label{sec: nearly timelike maximizers}

If $\g$ is a nearly timelike curve from $p$ to $q$ such that $L(\g) = \tau(p,q)$ where $\tau$ is the time separation function introduced in the previous section (defined with respect to nearly timelike curves), then we call $\g$ a \emph{nearly timelike maximizer} from $p$ to $q$. Note that $\tau(p,q) < \infty$ whenever a nearly timelike maximizer  from $p$ to $q$ exists. 

In this section we establish sufficient conditions ensuring the existence of a nearly timelike maximizer between two points $p$ and $q$ in a $C^0$ spacetime $(M,g)$, see Theorem \ref{thm: nearly timelike maximizer}. We adopt the notation
\[
\mc{J}(p,q) \,:=\, \mc{J}^+(p) \cap \mc{J}^-(q).
\]
Also, we need the following two definitions:
\begin{itemize}
\item[$\bullet$] $(M,g)$ is \emph{strongly causal} if for every $p \in M$ and every neighborhood $U$ of $p$, there is a neighborhood $V \subset U$ of $p$ such that 
\[
\g(a), \g(b) \in V \quad \Longrightarrow \quad \g \subset U
\]
whenever $\g \colon [a,b] \to M$ is a causal curve. 
\item[$\bullet$] A subset $E \subset M$ is called \emph{causally plain} if 
\[
q \in J^+(p)  \quad \Longrightarrow  \quad q \in \ov{I^+(p)} \:\: \text{ and } \:\: p \in \ov{I^-(q)}
\] 
for all $p,q \in E$.
\end{itemize}

%

\medskip

\noindent{\bf Remark 4.1.} An assumption used throughout this section is that $\mc{J}(p,q)$ is causally plain. We emphasize that this does \emph{not} imply that no bubbling from $p$ or $q$ exists, but it does imply the lack of bubbling from within $\mc{J}(p,q)$.  Proposition \ref{prop: summary} outlines the class of bubbling spacetimes to which the results in this section are relevant; a specific example based on the spacetime in \cite{Leonardo_Soultanis} is provided after that proposition. Also, we do not need the full strength of strong causality for the results in this section. It would be sufficient to consider strong causality on $\mc{J}(p,q)$, and, in fact, it would be sufficient to consider only nearly timelike curves instead of causal curves in the definition.

\medskip
\medskip

The following lemma establishes a limit curve argument for nearly timelike curves. Recall from the example after Corollary \ref{cor: cont spacetimes are LpreLS} that a limit curve argument for nearly timelike curves does not hold in general, so additional assumptions need to be imposed.

\medskip

\begin{customlemma}{4.2}\label{lem: limit curve}
Suppose $(M,g)$ is a strongly causal $C^0$ spacetime. Assume  $\mc{J}(p,q)$ is compact and causally plain. Let $\g_n \colon [0,b_n] \to M$ be a sequence of nearly timelike curves from $p_n$ to $q_n$ parameterized by $h$-arclength. Assume 
\[
 p_n \to p, \quad q_n \to q,  \quad p_n \in \mc{J}^+(p),  \quad q_n \in \mc{J}^-(q),\quad\text{ and } \quad p \neq q.
\]
Then there is a $b \in (0, \infty)$ and a nearly timelike  curve $\g\colon [0,b] \to M$ from $p$ to $q$ such that for each $t \in(0,b)$, there is a subsequence $\g_{n_k}$ which converges to $\g$ uniformly on $[0,t]$.
\end{customlemma}

\proof
We first show $\sup_{n}\{b_n\} < \infty$.   By assumption, $p_n \in \mc{J}^+(p)$. Therefore $\g_n(t) \in \mc{J}^+(p)$ by Theorem \ref{thm: lsc}(2) for each $t \in [0,b_n]$. Likewise $\g_n(t) \in \mc{J}^-(q)$. Therefore each $\g_n$ is contained in the compact set $\mc{J}(p,q)$. By \cite[Prop. 2.17]{Ling_causal_theory}, for each $x \in\mc{J}(p,q)$, there is a neighborhood $U_x$ such that $L_h(\l) \leq 1$ for any causal (and hence any nearly timelike) curve $ \l \subset U_x$. By strong causality, there are  neighborhoods $V_x \subset U_x$ such that $\l \subset U_x$ whenever $\l \colon [a,b] \to M$ is a causal  curve with endpoints in $V_x$.  Since $\mc{J}(p,q)$ is covered by $\{V_x\}_{x \in \mc{J}(p,q)}$, there is a finite subcover $V_1, \dotsc, V_N$. It follows that any nearly timelike curve with image contained in $\mc{J}(p,q)$ has $h$-length bounded by $N$. Thus $\sup_{n}\{b_n\} \leq N$.

Since every sequence in $\R$ contains a monotone subsequence, we can assume $b_n$ is monotone by restricting to a subsequence. Then either (1) $b_n \to \infty$ or (2) $b_n \to b < \infty$. The first scenario is ruled out by the paragraph above. Therefore the second scenario must hold. Moreover, $b > 0$. Indeed, we have $d_h(p_n,q_n) \leq b_n$, and taking $n \to \infty$ gives $d_h(p,q) \leq b$. Thus the assumption $p \neq q$ implies $b > 0$.

Extend each $\g_n$ to inextendible causal curves $\tilde{\g}_n \colon \R \to M$ by, for example, concatenating each $\g_n$ with the maximal integral curve of a timelike vector field and then reparamterizing by $h$-arclength. By the usual limit curve theorem \cite[Thm. 2.21]{Ling_causal_theory}, there exists a subsequence (still denoted by $\tilde{\g}_n$) and a causal curve $\tilde{\g}\colon \R \to M$ with $\tilde{\g}(0) = p$ such that $\tilde{\g}_n$ converges to $\tilde{\g}$ uniformly on compact subsets of $\R$. The triangle inequality gives
\begin{align*}
d_h\big(q, \tilde{\g}_{n}(b)\big) \,&\leq\, d_h\big(q, \g_{n}(b_{n})\big) \,+\, d_h\big(\g_{n}(b_{n}), \tilde{\g}_{n}(b)\big) 
\\
&\leq\, d_h\big(q, \g_{n}(b_{n})\big) \,+\, |b_n-b|.
\end{align*}
Since $\g_{n}(b_{n}) \to q$ and $b_{n} \to b$, the right-hand side limits to 0. Thus $\tilde{\g}_{n}(b) \to q$.  Therefore $\tilde{\g}|_{[0,\, b]}$ is a causal curve from $p$ to $q$.  Set $\g = \tilde{\g}|_{[0,b]}$. Fix $t \in (0,b)$. There is a subsequence (still denoted by $b_n$) such that $b_n \geq t$ for all $n$. Therefore, for this subequence, we have $\g_n = \tilde{\g}_n$ on $[0,t]$; hence $\g_n$ converges uniformly to $\g$ on $[0,t]$.

It remains to show that $\g$ is a nearly timelike curve. Fix $s < t$ in $[0,b]$. By compactness, we have $\g \subset \mc{J}(p,q)$. Therefore $\g(t) \in \ov{I^+\big(\g(s)\big)}$ since $\mc{J}(p,q)$ is causally plain. Likewise $\g(s) \in \ov{I^-\big(\g(t)\big)}$. \qed

\medskip
\medskip

\begin{customthm}{4.3}\label{thm: nearly timelike maximizer}
Suppose $(M,g)$ is a strong causal $C^0$ spacetime. Assume $\mc{J}(p,q)$ is compact and causally plain. If $q \in \mc{J}^+(p)$ with $q \neq p$, then there is a nearly timelike maximizer $\g$ from $p$ to $q$, i.e., $L(\g) = \tau(p,q)$.
\end{customthm}

\proof
By definition of $\tau$, there is a sequence of nearly timelike curves $\g_n \colon [0,b_n] \to M$ from $p$ to $q$ satisfying $\tau(p,q) \leq L(\g_n) + 1/n$. Assume each $\g_n$ is parameterized by $h$-arclength. Let $\g \colon [0, b] \to M$ be the nearly timelike curve from $p$ to $q$ appearing in the conclusion of Lemma \ref{lem: limit curve}. As in the proof of that lemma, let $\tilde{\g}_n \colon \R \to M$ be the inextendible causal curve extensions of $\g_n$ and let $\tilde{\g}\colon \R \to M$ be the resulting limit curve so that $\g = \tilde{\g}|_{[0,b]}$.

It suffices to show $L(\g) \geq \tau(p,q)$. There is a subsequence $\tilde{\g}_{n_k}|_{[0,b]}$ which converges uniformly to $\g$; moreover, $b_{n_k}$ limits to $b$ monotonically as $k \to \infty$. Fix $\e > 0$. By upper semicontinuity of the Lorentzian length functional \cite[Prop. 3.7]{Ling_causal_theory}, there is an $N$ such that $k \geq N$ implies 
\begin{align*}
L(\g) + \e \,&\geq\, L\big(\tilde{\g}_{n_k}|_{[0,b]}\big)
\\
&=\, L(\g_{n_k}) + \int_{b_{n_k}}^b \sqrt{-g\big(\tilde{\g}_{n_k}',\tilde{\g}_{n_k}'\big)}
\\
&\geq\, \big(\tau(p,q) - 1/n_k\big) + \int_{b_{n_k}}^b\sqrt{-g\big(\tilde{\g}_{n_k}',\tilde{\g}_{n_k}'\big)}.
\end{align*}
As $k \to \infty$, the above integral limits to zero. This follows since (1) $b_{n_k} \to b$ and (2) there is a neighborhood $U$ of $q$ such that $-g(\l',\l')$ is bounded on $U$ for any $h$-arclength parmeterized curve $\l$ contained in $U$. (1) is clear. To prove (2), let $U$ be a coordinate neighborhood of $q$ with coordinates $x^\mu$, and assume $U$ is $h$-convex and has compact closure. Using similar triangle inequality arguments as in the proof of \cite[Prop. 2.2]{Ling_causal_theory}, it follows that the component functions $\l^\mu = x^\mu \circ \l$ of any $h$-arclength parameterized curve $\l$ are Lipschitz with the same Lipschitz constant; this proves (2). Thus, taking $k \to \infty$, we have $L(\g) + \e \geq \tau(p,q)$. Since $\e > 0$ was arbitrary, we have $L(\g) \geq \tau(p,q)$.  
\qed

%

\medskip
\medskip


The next theorem proves a sequential continuity result for $\tau$ but only from directions within $\mc{J}(p,q)$ and under the assumption that $\mc{J}(p,q)$ is compact and causally plain. It could have applications to synthetic approaches of Lorentzian geometry which require a continuous $\tau$ as in \cite{BLMS}.

\medskip

\begin{customthm}{4.4}\label{thm: usc}
Suppose $(M,g)$ is a strongly causal $C^0$ spacetime. Assume $\mc{J}(p,q)$ is compact and causally plain.
If
\[
 p_n \to p, \quad q_n \to q,  \quad p_n \in \mc{J}^+(p),  \quad q_n \in \mc{J}^-(q),\quad\text{ and } \quad p \neq q,
\]
then 
\[
\lim_{n \to \infty} \tau (p_n, q_n) \,=\, \tau(p,q).
\]
\end{customthm}

\proof
By lower semicontinuity of $\tau$, we have $\tau(p,q) \leq \liminf \tau(p_n, q_n)$. It suffices to show $\tau(p,q) \geq \limsup \tau(p_n, q_n)$. Set $t := \limsup \tau(p_n, q_n)$. If $t = 0$, then $\tau(p,q) \geq t$ is immediate. Therefore we can assume $t > 0$. Seeking a contradiction, suppose $\tau(p,q) < t$. Then there are subsequences (still denoted by $p_n$ and $q_n$) and an $\e > 0$ such that $\tau(p,q) < \tau(p_{n}, q_{n}) -2\e$ for all $n$ and $\tau(p_{n}, q_{n}) \to t$ as $n \to \infty$. Since $t > 0$, we can assume $\tau(p_{n}, q_{n}) > 0$ for all $n$ by restricting to a further subsequence. Let $\g_n \colon [0,b_n] \to M$ be a sequence of nearly timelike curves from $p_{n}$ to $q_{n}$ such that $L(\g_{n}) > \tau(p_{n}, q_{n}) - 1/n$. Let $\g \colon [0, b] \to M$ be the nearly timelike curve from $p$ to $q$ appearing in the conclusion of Lemma \ref{lem: limit curve}. As in the proof of that lemma, let $\tilde{\g}_n \colon \R \to M$ be the inextendible causal curve extensions of $\g_n$ and let $\tilde{\g}\colon \R \to M$ be the resulting limit curve so that $\g = \tilde{\g}|_{[0,b]}$. There is a subsequence $\tilde{\g}_{n_k}|_{[0,b]}$ which converges uniformly to $\g$; moreover, $b_{n_k}$ limits to $b$ monotonically as $k \to \infty$. By upper semicontinuity of the Lorentzian length functional \cite[Prop. 3.7]{Ling_causal_theory}, there is an $N$ such that $k \geq N$ implies 
\begin{align*}
L(\g) + \e \,&\geq\, L\big(\tilde{\g}_{n_k}|_{[0,b]}\big)
\\
&=\, L(\g_{n_k}) + \int_{b_{n_k}}^b \sqrt{-g\big(\tilde{\g}_{n_k}',\tilde{\g}_{n_k}'\big)}
\\
&>\, \big(\tau(p_{n_k},q_{n_k}) - 1/n_k\big) + \int_{b_{n_k}}^b\sqrt{-g\big(\tilde{\g}_{n_k}',\tilde{\g}_{n_k}'\big)}
\\
&>\, \big(\tau(p,q) + 2\e - 1/n_k\big) + \int_{b_{n_k}}^b\sqrt{-g\big(\tilde{\g}_{n_k}',\tilde{\g}_{n_k}'\big)}.
\end{align*}
As in the proof of Theorem \ref{thm: nearly timelike maximizer}, the integral term vanishes as $k \to \infty$. Therefore, we obtain $L(\g) \geq \tau(p,q) + \e$, which is a contradiction.
\qed

%

\medskip
\medskip

The previous results rely on $\mc{J}(p,q)$ being compact and causally plain. 
The following proposition gives sufficient conditions ensuring this and summarizes the results in this section. A specific example for which the proposition applies follows afterwards. Recall that a $C^0$ spacetime $(M,g)$ is \emph{globally hyperbolic} if it is strongly causal and $J^+(p) \cap J^-(q)$ is compact for all $p,q \in M$.

\medskip
\begin{customprop}{4.5}\label{prop: summary}
Let $(M,g)$ be a globally hyperbolic $C^0$ spacetime.  For an open subset $M' \subset M$, assume $g$ is smooth on $M'$ (locally Lipschitz is sufficient) and that $J^+(M') \subset M'$.  If $\ov{I^+(p)} \setminus \{p\} \subset M'$ for some $p \in \ov{M'}$, then for all $q \in \mc{J}^+(p)$ with $q \neq p$, the following hold:
\begin{itemize}
\item[\emph{(1)}] $\mc{J}(p,q)$ is compact and causally plain.
\item[\emph{(2)}] There is a nearly timelike maximizer from $p$ to $q$.
\item[\emph{(3)}] $\displaystyle\lim_{n \to \infty} \tau(p_n, q_n) = \tau(p,q)$ if $p_n \to p$ and $q_n \to q$ with $p_n \in \mc{J}^+(p)$ and $q_n \in \mc{J}^-(q)$. 
\end{itemize}
\end{customprop}

\proof
(2) and (3) follow from (1) via Theorems \ref{thm: nearly timelike maximizer} and \ref{thm: usc}, respectively, so it suffices to show (1). We first show that $\mc{J}(p,q)$ is causally plain. Fix $x,y \in \mc{J}(p,q)$ with $y \in J^+(x)$. We want to show $y \in \ov{I^+(x)}$ and $x \in \ov{I^-(y)}$. Either $x = p$ or $x \neq p$. If $x = p$, then $y \in \mc{J}^+(x)$ and so the result follows. Now assume $x \neq p$. Then $x \in \ov{I^+(p)} \setminus \{p\} \subset M'$. Therefore any causal curve from $x$ to $y$ will be contained in $M'$ since $J^+(M') \subset M'$. The metric is smooth on $M'$ and so the push-up property holds on $M'$ from which the result follows. (In fact this argument proves that $\mc{J}^+(p)$ is causally plain.)

 Now we show $\mc{J}(p,q)$ is compact. First note that $J(p,q) := J^+(p) \cap J^-(q)$ is compact by global hyperbolicity. By the Hopf-Rinow theorem, $J(p,q)$ is closed and bounded (with respect to the Riemannian distance function $d_h$). Since $\mc{J}(p,q) \subset J(p,q)$, it follows that $\mc{J}(p,q)$ is also bounded. Therefore it suffices to show that $\mc{J}(p,q)$ is closed. Let $r$ be a limit point of $\mc{J}(p,q)$. We can assume $r \neq p,q$. Let $r_n \in \mc{J}(p,q)$ be a sequence with $r_n \to r$. Let $\g_n \colon [0,b_n] \to M$ be a sequence of $h$-arclength parameterized nearly timelike curves from $p$ to $r_n$. Since $J(p,q)$ is compact, \cite[Prop. 3.4]{Ling_causal_theory} and its proof imply that there is a $b \in (0,\infty)$ with $b_n \to \infty$ and a causal curve $\g \colon [0, b] \to M$ from $p$ to $r$ such that for any $t \in (0,b)$, there is a subsequence of $\g_n$ which converges to $\g$ uniformly on $[0,t]$.

\medskip

\noindent Claim: $\g$ is a nearly timelike curve from $p$ to $r$.

\medskip

\noindent To prove the claim, the following will be useful. Fact: $\g(t) \in M'$ for all $t \in (0,b]$.

\medskip 

We first prove the fact. It suffices to show $\g(t) \in \ov{I^+(p)} \setminus \{p\}$.  $\g(t) \neq p$ since there are no closed causal curves in $M$.  If $t < b$, then there is a subsequence $\g_n(t) \to \g(t)$. If $t = b$, then $r_n \to \g(b)$. 

Now we prove the claim. Fix $s < t$ in $[0,b]$. We want to show $\g(t) \in \ov{I^+\big(\g(s)\big)}$ and $\g(s) \in \ov{I^-\big(\g(t)\big)}$. Either $s = 0$ or $s \neq 0$. Consider first $s \neq 0$. Then $\g(s), \g(t) \in M'$, by the fact, and so the result follows since the push-up property applies on $M'$. Now assume $s = 0$. The proof of the fact shows that $\g(t) \in \ov{I^+(p)}$.
 Now we show $p \in \ov{I^-\big(\g(t)\big)}$. If $U$ is any neighborhood of $p$, choose $0 < \e < t$ small enough so that $\g(\e) \in U$. The fact implies $\g(\e) \in \ov{I^-\big(\g(t)\big)}$, hence $U$ intersects $I^-\big(\g(t)\big)$. This proves the claim. 
 
 Thus $r \in \mc{J}^+(p)$. An easier argument shows that  $r \in \mc{J}^-(q)$. \qed

\medskip
\medskip

\noindent {\bf Example 4.6.} Let $(M,g)$ be  the Garc{\'i}a-Heveling-Soultanis $C^0$ spacetime \cite{Leonardo_Soultanis} which is globally hyperbolic. Let $P$ denote the null cone $t = |x|$. The metric is smooth on $M' := I^+(P)$, and since the lightcones of $g$ are narrower than those for the Minkowski metric, it follows that $J^+(M') \subset M'$. Moreover, for each $p \in P$, direct calculations as in \cite{Leonardo_Soultanis} show that $\ov{I^+(p)} \setminus \{p\} \subset M'$.  Therefore the previous proposition applies. In particular, for any $p \in P$ and $q \in \mc{J}^+(p)$ with $q \neq p$, there is a nearly timelike maximizer from $p$ to $q$. This spacetime also demonstrates that global hyperbolicitiy does not necessarily imply compactness of $\mc{J}(p,q)$ for all $p,q$. In $(t,x)$ coordinates, let $p = (-1,1)$ and $q = (1,1)$.  Then $\mc{J}(p,q)$ is not compact since it does not contain the origin $0 = (0,0)$ which is a limit point of $\mc{J}(p,q)$. See Figure \ref{fig: 4}.

\begin{figure}[h]
\[
\begin{tikzpicture}[scale = 1]

\draw [white, thick, fill = gray] (2,-4) -- (0,-2) -- (2,0) -- (4,-2) -- (2,-4);

\draw [thick] (0,-2) -- (2,-4) -- (4,-2) -- (2,0);

\draw [thick] (2,0) -- (4.1,2.1);
\draw [thick] (0,-2) -- (-4.1,2.1);

\draw [thick, white] (0,-2) -- (2,0);

\draw [thick, dashed] (0,-2) -- (2,0);

\draw [->] [thick] (-3.75,-0.5) arc [start angle=-90, end angle=-30, radius=40pt];
\draw (-4.5,-0.5) node [scale = .85] {$\pd J^+(0)$};

\draw [->] [thick] (-2,2.5) arc [start angle=10, end angle=-50, radius=35pt];
\draw (-2,2.9) node [scale = .85] {$\pd I^+(0)$};

\draw [<->,thick] (0,-4.75) -- (0,2.35);
\draw [<->,thick] (-4.5,-2) -- (4.5,-2);

\draw (-.35,2.5) node [scale = .85] {$t$};
\draw (4.75, -2.25) node [scale = .85] {$x$};

\draw [thick, blue] (0,-2) .. controls (2.5,1) .. (3,2.1);
\draw [thick, blue] (0,-2) .. controls (-2.5,1) .. (-3,2.1);

\node [scale = .50] [circle, draw, fill = black] at (2,0)  {};
\draw (2.5,0) node [scale = .85] {\small{$q$}};

\node [scale = .50] [circle, draw, fill = black] at (2,-4)  {};
\draw (1.5,-4.1) node [scale = .85] {\small{$p$}};

\node [scale = .50] [circle, draw, fill = white] at (0,-2)  {};
\draw (-.3,-2.3) node [scale = .85] {\small{$0$}};

\end{tikzpicture}
\]
\caption{\small{The Garc{\'i}a-Heveling-Soultanis $C^0$ spacetime $(M,g)$ is globally hyperbolic, and so the diamond $J(p,q)$ is compact. But this does not imply compactness of $\mc{J}(p,q)$. }}
\label{fig: 4}
\end{figure}

\section{Discussion and conclusion}\label{sec: conclusion}

In this paper, we defined a new class of curves for $C^0$ spacetimes dubbed ``nearly timelike curves" and introduced the relation $\mc{J}$ via $q \in \mc{J}^+(p)$ if there is a nearly timelike curve from $p$ to $q$ or if $q = p$. It satisfies $I^+(p) \subset \mc{J}^+(p) \subset J^+(p)$.  The motivation for introducing $\mc{J}$ was to obtain a lower semicontinuous time separation function $\tau$ (Theorem \ref{thm: lsc}). Therefore $C^0$ spacetimes can fit into the framework of Lorentzian pre-length spaces introduced in \cite{KS} (Corollary \ref{cor: cont spacetimes are LpreLS}). However, as demonstrated at the end of section \ref{sec: nearly timelike curves}, the Lorentzian pre-length space associated with a $C^0$ spacetime is not necessarily locally causally closed, which is an extra axiom in the definition of a Lorentzian length space. In summary, $C^0$ spacetimes fit into the frame work of Lorentzian pre-length spaces but not necessarily Lorentzian length spaces.

In section \ref{sec: nearly timelike maximizers}, we found sufficient conditions guaranteeing a limit curve argument for nearly timelike curves (Lemma \ref{lem: limit curve}) and also the existence of a nearly timelike maximizer between two points (Theorem \ref{thm: nearly timelike maximizer}), i.e., a nearly timelike curve $\g$ from $p$ to $q$ such that $L(\g) = \tau(p,q)$. We found sufficient conditions guaranteeing a sequential continuity  result, $\tau(p_n, q_n) \to \tau(p,q)$ (Theorem \ref{thm: usc}), but we must assume that the sequences $p_n \to p$ and $q_n \to q$ are contained in $\mc{J}^+(p)$ and $\mc{J}^-(q)$, respectively. This may have applications for the bounded Lorentzian metric-spaces introduced in \cite{BLMS} since they require a continuous time separation function. Our results are applied to a class of spacetimes in Proposition \ref{prop: summary} and a specific example is provided afterwards.


One can define \emph{$\mc{J}$-global hyperbolocity} by strong causality and compactness of all diamonds $\mc{J}(p,q)$.  Since $\mc{J}$-global hyperbolocity is not implied by the usual notion of global hyperbolocity  (see Figure \ref{fig: 4}), it would be interesting to characterize the notion of $\mc{J}$-global hyperbolocity. For example, is there an analogous concept of a Cauchy surface? In fact, the hierarchy of the causal ladder is not well understood for $C^0$ spacetimes, and so a detailed description of this hierarchy and how $\mc{J}$-global hyperbolocity fits into it would be an interesting question. 

  Other relations satisfying the push-up property have been defined for $C^0$ spacetimes as in \cite{Leonardo}. It would be interesting to compare the relationship between nearly timelike curves and $\tilde{d}^+$-curves in \cite{Leonardo}.   
   Also, an interesting question would be to determine if nearly timelike maximizers have a causal character, which is the case for causal maximizers in locally Lipschitz spacetimes \cite{GrafLing}. See also \cite{Lorentz_meets_Lipschitz, gannon_lee_note} for related results.

\section*{Acknowledgments} 
We thank Argam Ohanyan for bringing this question to our attention when he was visiting the University of Copenhagen and for helpful comments on a first draft. We thank Leonardo Garc{\' i}a-Heveling for clarification on the example in \cite{Leonardo_Soultanis}. We also thank Ettore Minguzzi, Stefan Suhr, Clemens S{\"a}mann,  Michael Kunzinger, and Greg Galloway for their valuable comments and suggestions. This work was supported by Carlsberg Foundation CF21-0680 and Danmarks Grundforskningsfond CPH-GEOTOP-DNRF151.

\appendix

\section{Relating the different definitions for timelike curves}\label{appendix}

In this appendix, we review the different definitions of timelike curves used in low-regularity causal theory and the relationships between them.

\medskip

\begin{Def}\emph{Let $\g \colon [a,b] \to M$ be a causal curve from $p$ to $q$.
\begin{itemize}
\item[(1)] $\g$ is \emph{timelike} if there exists an $\e > 0$ such that $g(\g', \g') < -\e$ almost everywhere.
\item[(2)] $\g$ is \emph{almost everywhere timelike} if $g(\g', \g') < 0$ almost everywhere.
\item[(3)] $\g$ is \emph{piecewise $C^1$ timelike} if $\g$ is piecewise $C^1$ and $\g'(t)$ is future directed timelike for all $t$ including the finite number of break points (understood as one-sided limits). 
\item[(4)] $\g$ is \emph{locally uniformly timelike} if there is a smooth Lorentzian metric $\check{g}$ such that $\check{g} < g$ and $\check{g}(\g', \g') < 0$ almost everywhere. ($\check{g} < g$ means ``$\check{g}(X,X) \leq 0$ implies $g(X,X) < 0$ for all nonzero $X$.")
\end{itemize}
}
\end{Def}

\medskip

\noindent The corresponding timelike futures are:

\begin{itemize}
\item[(1)] $I^+(p) = \{q \mid \text{there is a timelike curve from $p$ to $q$}\}.$

\item[(2)] $I^+_{\rm a.e.}(p) = \{q \mid \text{there is an almost everywhere timelike curve from $p$ to $q$}\}.$

\item[(3)] $I^+_{C^1}(p) = \{q \mid \text{there is a piecewise $C^1$ timelike curve from $p$ to $q$}\}.$

\item[(4)] $\check{I}^+(p) = \{q \mid \text{there is a locally uniformly timelike curve from $p$ to $q$}\}.$
\end{itemize}

\medskip

Definition (1) was introduced in \cite{Ling_causal_theory};  it's used in this paper and in \cite{Ling_remarks_cosmo,Greg_Graf_Ling_AdSxS2,B&B2}. A proof showing openness of $I^+(p)$ is given in  \cite[Thm. 2.12]{Ling_causal_theory}. Definition (2) is not used as much since $I^+_{\rm a.e.}(p)$ is not necessarily open, see \cite[Ex. 3.1]{future_not_open}; however, if the metric is smooth (locally Lipschitz is sufficient), then $I^+_{\rm a.e.}(p) = I^+(p)$, see Proposition \ref{prop: timelike futures} below. Definition (3) is the most widely used, e.g., \cite{SbierskiHol, SbierskiSchwarz1, SbierskiSchwarz2, GalLing_con, GLS,Ling_coord_sing,GJL,B&B,Sbierski_FLRW, Ling_C0_FLRW}. A proof showing openness of $I^+_{C^1}(p)$ is given in \cite[Prop. 2.6]{SbierskiSchwarz1}, see also \cite[Prop. 2.2]{Ling_coord_sing}.  Definition (4) was introduced in \cite{ChrusGrant}. That $\check{I}^+(p)$ is open is shown in \cite[Prop. 1.4]{ChrusGrant}.

 The following proposition shows that definitions (1), (3), and (4) agree with each other. It's a mild generalization of \cite[Lem. 2.7]{future_not_open}.

\medskip

\begin{prop}\label{prop: timelike futures}
For all points $p$ in a $C^0$ spacetime $(M,g)$, we have
\[
I^+(p) \,=\, \check{I}^+(p) \,=\, I^+_{C^1}(p).
\]
Moreover, if $g$ is smooth (locally Lipschitz is sufficient), then they're all equal to $I^+_{\rm a.e.}(p)$.
\end{prop}

\proof
First recognize that 
\[
\check{I}^+(p) \,\subset\, I^+_{C^1}(p) \,\subset\, I^+(p).
\]
The first inclusion follows from causal theory for smooth (at least $C^2$) metrics: the endpoints of an almost everywhere timelike curve can be joined by a broken timelike geodesic in a smooth spacetime, see \cite[Cor. 2.4.11]{ChrusBHbook}. The second inclusion follows from compactness, see \cite[Prop. 2.4]{Ling_causal_theory}. Conversely, fix $q \in I^+(p)$. It suffices to show $q \in \check{I}^+(p)$. Let $\g \colon [0,b] \to M$ be a timelike curve from $p$ to $q$ and assume $\g$ is parameterized by $h$-arclength. Then there is an $\e>0$ such that $g(\g', \g') < -\e$ almost everywhere. By \cite[Prop. 1.2]{ChrusGrant}, there is a smooth Lorentzian metric $\check{g}$ satisfying $\check{g} < g$ and $d(\check{g}, g) < \e$ where 
\[
d(\check{g}, g) \,=\, \sup_{p \in M,\: 0\neq X,Y \in T_pM}\frac{|\check{g}(X,Y) - g(X,Y)|}{|X|_h|Y|_h}.
\]
Since $\g$ is parameterized by $h$-arclength, $d(\check{g}, g) < \e$ implies $\check{g}(\g', \g')  <\e + g(\g', \g') < 0$ almost everywhere. Thus $q \in \check{I}^+(p)$.

Lastly, suppose $g$ is smooth. Then the endpoints of an almost everywhere timelike curve can be joined by a broken timelike geodesic, hence $I^+_{\rm a.e.}(p) \subset I^+_{C^1}(p)$. Therefore the four timelike futures are equal in this case. That locally Lipschitz is sufficient follows from Corollary 1.17 and Proposition 1.21 in \cite{ChrusGrant}.
\qed

%

\bibliographystyle{amsplain}

\end{document}